\documentclass[twocolumn,aps,prd,showpacs,nofootinbib]{revtex4}
\usepackage[russian]{babel}
\usepackage{graphics,graphicx}
\usepackage{amssymb,amsfonts}

\def\nq{\hspace*{-1em}}          
       \def\nhh{\hspace*{-0.3em}}
\def\cm{\hspace*{1cm}}

\def\ten#1{\mbox{$\ \cdot\ 10^{#1}$}}

\def\Jl#1#2{#1 { \bf #2},\ }

\def\ApJ#1 {\Jl{Astroph. J.}{#1}}
\def\CQG#1 {\Jl{Class. Quantum Grav.}{#1}}
\def\DAN#1 {\Jl{Dokl. AN SSSR}{#1}}
\def\GC#1 {\Jl{Grav. Cosmol.}{#1}}
\def\GRG#1 {\Jl{Gen. Rel. Grav.}{#1}}
\def\JETF#1 {\Jl{Zh. Eksp. Teor. Fiz.}{#1}}
\def\JETP#1 {\Jl{Sov. Phys. JETP}{#1}}
\def\JHEP#1 {\Jl{JHEP}{#1}}
\def\JMP#1 {\Jl{J. Math. Phys.}{#1}}
\def\NPB#1 {\Jl{Nucl. Phys.}{B\ #1}}
\def\NP#1 {\Jl{Nucl. Phys.}{#1}}
\def\PLA#1 {\Jl{Phys. Lett.}{#1A}}
\def\PLB#1 {\Jl{Phys. Lett.}{#1B}}
\def\PRD#1 {\Jl{Phys. Rev.}{D\ #1}}
\def\PRL#1 {\Jl{Phys. Rev. Lett.}{#1}}

\mathsurround 1pt                      \def\lal{&& {}\nhh}
\def\eq{Eq.\,}                         \def\eqs{Eqs.\,}
\def\beq{\begin{equation}}             \def\eeq{\end{equation}}
\def\bear{\begin{eqnarray}}            \def\bearr{\begin{eqnarray} \lal}
\def\ear{\end{eqnarray}}               \def\earn{\nonumber \end{eqnarray}}
\def\nn{\nonumber\\ {}}                \def\nnv{\nonumber\\[5pt] {}}
\def\nnn{\nonumber\\ \lal }            \def\nnnv{\nonumber\\[5pt] \lal }
                    \def\yyy{\\[5pt] \lal }
\def\eql{&\! = &\!}                    

\def\dst{\displaystyle}                
\def\fracd#1#2{{\dst\frac{#1}{#2}}}    
\def\Half{{\fracd{1}{2}}}              

        \def\e{{\,\rm e}}
\def\d{\partial}                       
      
      \def\sign{\mathop{\rm sign}\nolimits}
  \def\dim{\mathop{\rm dim}\nolimits}
\def\const{{\rm const}}                \def\eps{\varepsilon}

\def\then{\ \Rightarrow\ }

\def\DAL{\mathop{\raisebox{-.5pt}{$\Box$}}\nolimits}

                 \def\wt{\widetilde}
                 \def\tg{{\wt g}}
\def\mn{_{\mu\nu}}                  \def\tR{{\wt R}}

                   \def\M{{\mathbb M}}
\def\cK{{\cal K}}                   
\def\cV{{\cal V}}                   
            
\def\oR{{\overline R}}              
             
\def\kappa{\varkappa}

\def\Lambdaef{\Lambda_{\rm eff}}

        \def\wt{\widetilde}
            \def\tg{{\wt g}}
\def\mD{m_{\sss D}}                 \def\tR{{\wt R}}
                    
\def\mD{m_{{}_{D}}}
               \def\ov{\overline}
                    \def\oR{\ov{R}}
\def\cK{{\cal K}}                   
\def\cV{{\cal V}}                   
\def\eff{_{\rm eff}}

\def\sumi{\sum\nolimits_i}

\begin{document}

\title{Multidimensional world, inflation and modern acceleration}

\author{K.A. Bronnikov}
\affiliation
    {Center for Gravitation and Fundamental Metrology, VNIIMS,
            46 Ozyornaya St., Moscow 119361, Russia;
     Institute of Gravitation and Cosmology, PFUR,
            6 Miklukho-Maklaya St., Moscow 117198, Russia
     \email {kb20@yandex.ru}}

\author{S.G. Rubin}
\affiliation
    {National Research Nuclear University ``MEPhI'',
        31 Kashirskoe Sh., Moscow 115409, Russia
     \email {sergeirubin@list.ru}}

\author{I.V. Svadkovsky}
\affiliation
    {National Research Nuclear University ``MEPhI'', 31 Kashirskoe Sh.,
        Moscow 115409, Russia}

\begin{abstract}
   Starting from pure multidimensional gravity with curvature-nonlinear terms
   but no matter fields in the initial action, we obtain a cosmological
   model with two effective scalar fields related to the size of two extra
   factor spaces. The model includes both an early inflationary stage and
   that of modern accelerated expansion and satisfies the observational data.
   There are no small parameters; the effective inflaton mass depends on the
   initial conditions which explain its small value as compared to the Planck
   mass. At the modern stage, the size of extra dimensions slowly increases,
   therefore this model predicts drastic changes in the physical laws of our
   Universe in the remote future.

\pacs{04.50.+h; 98.80.-k; 98.80.Cq}
\end{abstract}

\maketitle

\section{Introduction}

  The existence of an early inflationary stage has become a conventional
  feature in modern descriptions of the Universe due to great success of
  inflationary scenarios in explaining the observational data (see, e.g.,
  \cite{BR-book,bp08} for recent reviews). A great number of inflationary
  scenarios have been suggested by now, and this number is still rapidly
  growing. It is really difficult to single out a scenario that has been
  indeed realized by Nature. Another problem is related to the origin of the
  scalar field (or fields), the so-called inflaton(s), which are almost
  inevitable ingredients of such scenarios.

  On the other hand, the most important set of problems in modern cosmology
  are related to the observed accelerated expansion of the Universe. Its most
  popular explanation, fitting all observational constraints, is the
  so-called $\Lambda$CDM model, invoking a cosmological constant $\Lambda$ as
  a material source that causes the accelerated expansion via the Einstein
  equations \cite{LCDM}. However, the hardest problem of this model is the
  extremely small observed value of $\Lambda$ (usually ascribed to the
  physical vacuum density) as compared to the Planck density, the natural
  vacuum energy density of quantum fields: the corresponding ratio is about
  $10^{-123}$.

  Of greatest interest are scenarios that try to jointly describe the entire
  history of our Universe or at least such its important stages as the early
  inflation and the modern acceleration. A promising approach on this trend
  is to use modified theories of gravity, e.g., multidimensional ones. In
  our view, curvature-nonlinear multidimensional gravity is a good candidate.

  It has been recently argued \cite{VfromD,brost,asym_bw,as-accel} that
  multidimensional gravity with curvature-nonlinear terms in the action can
  be a source of a great diversity of effective theories able to address a
  number of important problems of modern astrophysics and cosmology using a
  minimal set of postulates. Among such problems one can mention the essence
  of dark energy, early formation of supermassive black holes (which is a
  necessary stage in some scenarios of cosmic structure formation), and
  sufficient particle production at the end of inflation. In this approach,
it is supposed that essentially different classical universes emerge
  from space-time foam due to quantum fluctuations, so that particular values
  of the total space-time dimension $D > 4$ and the topological properties of
  space-time may vary from one space-time region to another. Different
  effective theories can take place even with fixed parameters of the
  original Lagrangian. It can be shown that this approach, without need for
  fields other than gravity, is able to produce such different structures as
  inflationary (or simply accelerating) universes, brane worlds
  \cite{asym_bw}, black holes etc. The role of scalar fields is played by the
  metric components of extra dimensions.

  In the present paper, we show how pure nonlinear multidimensional gravity,
  without invoking any material source, makes it possible to describe, in a
  single scenario, an inflationary stage of the early Universe and a late
  accelerating stage with a sufficiently small effective cosmological
  constant. The model obtained agrees with the observational data.

  Let us mention some other approaches to obtaining such joint scenarios.
  Ref. \cite{NOd08} shows how to achieve this goal in some models of
  nonlocally modified gravity theories in four dimensions; in these models,
  the dark energy effect is caused by a composite graviton degree of freedom.
  In \cite{galtsov}, the same goal is achieved using a Yang-Mills condensate
  as a matter source. Ref. \cite{Gong} considers a relationship between
  hybrid inflation and dark energy; see there also numerous references on the
  subject.

  The paper is organized as follows. In Sec. II, we describe the general
  formalism. Sec. III shows how to obtain a successful inflationary scenario
  in a Kaluza-Klein type model with a single extra factor space. Sec. IV is
  devoted to obtaining models with two extra factor spaces able to unify
  inflation and modern acceleration. Sec. V is a conclusion.

\section{$D$-dimensional gravity}

  We will briefly describe a method of considering a wide classes of
  Lagrangians in multidimensional gravity in a Kaluza-Klein type approach,
  following \cite{VfromD,brost}. Consider the action\footnote
       {Our conventions are: the metric signature $(+{}-{}-{}\ldots)$; the
    curvature tensor
    $R^{\sigma}{}_{\mu\rho\nu} = \d_\nu\Gamma^{\sigma}_{\mu\rho}-\ldots,\
    R\mn = R^{\sigma}{}_{\mu\sigma\nu}$, so that the Ricci scalar
    $R > 0$ for de Sitter space-time and the matter-dominated
    cosmological epoch; the system of units $c = \hbar = 1$.}
\bearr                                      \label{S_D}
     S =  \Half \mD^{D-2} \int \sqrt{^D g}\,d^D x \biggl[ F(R)
        + c_1 R_{AB} R^{AB} + c_2 \cK \biggr]
\nnn
\ear
   in $D$-dimensional space-time $\M$ with the structure
$
     \M = \M_0 \times \M_1 \times \dots \times \M_n,
$
  where $\dim \M_i = d_i$ and $\mD$ is the $D$-dimensional Planck mass
  (not necessarily coinciding with the conventional Planck mass $m_4$),
  and the metric
\beq \label{ds_D}
    ds_D^2 = g_{ab}(x)dx^a dx^b
                    + \sum_{i=1}^{n}\e^{2\beta_i(x)} g^{(i)},
\eeq
  where $(x)$ means the dependence on the first $d_0$ coordinates $x^a$;
  $g_{ab} = g_{ab}(x)$ is the metric in $\M_0$, $g^{(i)}$ are $x$-independent
  $d_i$-dimensional metrics of the factor spaces $\M_i$, $i= \overline{1,n}$.
  In (\ref{S_D}), $F(R)$ is an arbitrary function of the scalar curvature $R$
  of $\M$; $c_1$ and $c_2$ are constants; $R_{AB}$ and $\cK= R_{ABCD}
  R^{ABCD}$ are the Ricci tensor and the Kretschmann scalar of $\M$,
  respectively; capital Latin indices cover all $D$ coordinates, small Latin
  ones ($a,b,\ldots$) the coordinates of $\M_0$, and $a_i,\ b_i, \ldots$ the
  coordinates of $\M_i$. Let us note that terms proportional to $R^2$ and
  other powers of $R$, $R_{AB}R^{AB}$ and the Kretschmann scalar $\cK =
  R_{ABCD}R^{ABCD}$ and other high-order curvature terms appear due to
  quantum corrections in quantum field theory in curved space-times
  \cite{GMM, BirD}.

  The $D$-dimensional Riemann tensor has the nonzero components
\bear
    R^{ab}{}_{cd} \eql \oR^{ab}{}_{cd},               \label{Riem_D}
\nn
    R^{a a_i}{}_{b b_i} \eql \delta^{a_i}_{b_i}\, B^a_b{}_{(i)}, \quad\
        B^a_b{}_{(i)} := \e^{-\beta_i} \nabla_b
            (\e^{\beta_i} \beta_i^{,a}),
\nn
    R^{a_i b_i}{}_{c_i d_i} \eql
      \e^{-2\beta_i} \oR^{a_i b_i}{}_{c_i d_i}
            + \delta^{a_i b_i}{}_{c_i d_i}
                \beta_{i,a}\beta_i^{,a},
\nn
    R^{a_i b_k}{c_i d_k} \eql
     \delta^{a_i}_{c_i} \delta^{b_k}_{d_k}
            \beta_{i,\mu}\beta_k^{,\mu},  \cm i\ne k.
\ear
  Here the bar marks quantities obtained from the factor space
  metrics $g_{ab}$ and $g^{(i)}$ taken separately,
  $\beta_{,a} \equiv \d_a \beta$, $\delta^{ab}{}_{cd}\equiv
  \delta_{c}^{a}\delta_{d}^{b}-\delta_{d}^{a}\delta_{c}^{b}$ and similarly
  for other kinds of indices.

  The nonzero components of the Ricci tensor and the scalar curvature are
\bear
    R_a^b \eql \oR_a^b + \sumi d_i\, B_a^b{}_{(i)},         \label{Ric_D}
\nn
    R_{a_i}^{b_i} \eql \e^{-2\beta_i} \oR_{a_i}^{b_i}
                    + \delta_{a_i}^{b_i}
                [ \DAL \beta_i + \beta_{i,a} \sigma^{,a}],
\nn
    R \eql \oR[g] + \sumi \e^{-2\beta } \oR_i + 2 \DAL \sigma
\nnn\cm \cm
        + (\d\sigma)^2 + \sumi d_i (\d \beta_i)^2,
\ear
  where $\sigma := \sum_i d_i \beta_i$; $(\d\sigma)^2 \equiv
  \sigma_{,a}\sigma^{,a}$ and similarly for other functions; $\DAL =
  g^{ab}\nabla_a \nabla_b$ is the $d_0$-dimensional d'Alembert operator;
  $\oR[g]$ and $\oR_i$ are the Ricci scalars corresponding to $g_{ab}$ and
  $g^{(i)}$, respectively. Here and henceforth $\sum_i$ means
  $\sum_{i=1}^{n}$.

\subsection* {Slow-change approximation. Reduction to lower dimensions}

  Let us suppose that all quantities are slowly varying, i.e., consider
  each derivative $\d_a$ (including those in the definition of $\oR$)
  as an expression containing a small parameter $\eps$, and neglect all
  quantities of orders higher than $O(\eps^2)$.
  Then we have the following decompositions:
\bear
       R \eql \phi + \oR[g] + f_1,
\nnn
      \ \ \ f_1 := 2\DAL \sigma + (\d\sigma)^2 + \sumi d_i (\d\beta_i)^2;
\nn
    F(R) \eql F (\phi) + F'(\phi)(\oR[g] + f_1) + O(\eps^4);
\nn
     R_{AB} R^{AB} \eql \sumi \frac{1}{d_i} \phi_i^2
\nnn
    + 2 \sumi d_i \phi_i [\DAL \beta_i + (\d\beta_i, \d\sigma)]+ O(\eps^4);
\nn
    \cK \eql 2\! \sumi \! \frac{\phi_i^2}{d_i(d_i{-}1)} +
            4 \sumi \! d_i \phi_i (\d\beta_i)^2  + O(\eps^4),
\nnn
\ear
  where
\beq                            \label{def-phi}
      \phi_i := K_i m_D^2 (d_i - 1) \e^{-2\beta_i}, \qquad
      \phi := \sumi d_i\phi_i.
\eeq
  The symbol $(\d\alpha, \d\beta)$ means $g^{ab}\alpha_{,a}\beta_{,b}$,
  and $F'(\phi) = dF/d\phi$.

  As a result, neglecting $o(\eps^2)$ and integrating out all $\M_i$, we
  obtain the following purely gravitational action reduced to $d_0$
  dimensions:
\bear
     S \eql \Half {\cV}\, \mD^d \!                           \label{SJ}
                \int \! \sqrt{g_0}\,d^{d_0} x\, \Big \{
                \e^{\sigma} F'(\phi)\oR_0
\nnn \
        + K_J \, -2 V_J(\phi_i) \Big\}
         + {\cV} \int \! \sqrt{g_0}\,d^{d_0} x\,\e^{\sigma} ,
\nnv
     K_J \eql  F'(\phi) \e^\sigma f_1
            +2\e^\sigma \sumi d_i\phi_i [c_1 \DAL\beta_i
\nnn \cm \
             + c_1 (\d\beta_i, \d\sigma) + 2c_2 (\d\beta_i)^2],
\nn
     -2V_J (\phi_i) \eql  \e^\sigma \Biggl[F(\phi)
     + \sumi d_i \phi_i^2 \biggl(c_1 + \frac{2c_2}{d_i-1}\biggr)\Biggr],
\nnn
\ear
  where $d = d_0-2$, $g_0 = |\det(g\mn)|$ and $\cV$ is a product of
  volumes of $n$ compact $d_i$-dimensional spaces $\M_i$ of unit
  curvature. The expression (\ref{SJ}) is typical of a
  (multi)scalar-tensor theory (STT) of gravity in a Jordan frame.

  Subtracting a full divergence, we get rid of second-order derivatives
  in (\ref{SJ}), and the resulting kinetic term takes the form
\bearr \label{KJ}   \nq
     K_J = F'\e^\sigma
            \biggl[-(\d\sigma)^2 + \sumi d_i(\d\beta_i)^2\biggr]
\nnn  \nq
    -2F'' \e^\sigma(\d\phi,\d\sigma)
            + 4\e^\sigma (c_1 + c_2)\sumi d_i \phi_i (\d\beta_i)^2,
\ear
  where $F'' = d^2 F/d\phi^2$.

\subsection* {Transition to the Einstein frame}

  For further analysis, it is helpful to pass on to the Einstein
  frame using the conformal mapping
\bearr \label{trans-g}
        g\mn \ \mapsto \tg\mn = |f(\phi_i)|^{2/(d_0-2)} g\mn,
\nnn \cm
       \cm   f(\phi_i) =  \e^{\sigma} F'(\phi).
\ear
  The expression with the scalar curvature in (\ref{SJ})
  transforms as follows:
\bearr \label{trans-R}
       \sqrt{g_0} \e^\sigma \oR_0 = \sqrt{g_0} f\oR_0
\nnn \ \ \
    = (\sign f) \sqrt{\tg} \biggl[
        \tR + \frac{d_0{-}1} {d_0{-}2}\ \frac{(\wt\d f)^2}{f^2} \biggr]
            + {\rm div},
\ear
  where the tilde marks quantities obtained from or with $\tg\mn$ and
  ${\rm div}$ denotes a full divergence which does not contribute to the
  field equations. The action (\ref{SJ}) acquires the form
\bearr                              \label{SE}
    S = \Half {\cV}\, \mD^d  \int \sqrt{\tg}\, d^{d_0} x
            \Bigl\{ [\sign F'(\phi)]\ \Big[\tR + K_E \Big]
\nnn \cm\cm
          - 2V_E (\phi_i)\Bigr\} 
\ear
  with the kinetic and potential terms
\bear
    K_E \eql \frac{1}{d} \biggl(\d\sigma
            + \frac{F''}{F'}\d\phi\biggr)^{\! 2}
                    + \biggl(\frac{F''}{F'}\biggr)^{\! 2} (\d\phi)^2
\nnn
         + \sumi d_i\biggl[ 1 + \frac{4}{F'} (c_1 {+} c_2)\phi_i\biggr]
              (\d\beta_i)^2,                                 \label{KE}
\\
     -2V_E (\phi_i) \eql                                     \label{VE}
            \e^{-2\sigma/d} |F'|^{-d_0/d}  \biggl[F(\phi)
\nnn \
      + \sumi d_i \phi_i^2
            \biggl(c_1  + \frac{2c_2}{d_i-1}\biggr)\biggr],
\ear
  where the tildes are omitted though the metric $\tg\mn$ is used, and
  $d: = d_0-2$; the indices are raised and lowered with $\tg\mn$. The
  original quantities $\beta_i$ and $\sigma$ are now expressed in terms of
  $n$ fields $\phi_i$ whose number coincides with the number of extra factor
  spaces.

  In what follows, we consider the most relevant case $d_0=4$ and
  accordingly $d=d_0-2 = 2$.

  A further interpretation of the results depends on which conformal frame
  is regarded physical (observational) \cite{bm-predict,erice}, and this in
  turn depends on the manner in which fermions appear in the (so far unknown)
  underlying unification theory involving all interactions. We will restrict
  ourselves to the simplest assumption, that the Einstein frame is
  simultaneously the observational frame. It means, in particular, that
  the effective Newtonian gravitational constant $G$ is a true constant in
  the course of the cosmological evolution. Moreover, we will assume for
  simplicity that the $D$-dimensional Planck mass $\mD$ is equal to the
  4-dimensional Planck mass $m_4 = 1/\sqrt{G}$ and, in what follows, we will
  put $G =1$, and numerical values of dimensionful parameters are thus
  expressed in Planck units.

\section{A single extra factor space and inflation}\label{single}

  Now, our program is as follows:
\begin{enumerate}
\item
    Choose the parameters of the original action (\ref{S_D}) to obtain a
    behavior of the potential (\ref{VE}) providing primordial inflation.
\item
    Additionally vary the parameters to satisfy the inflationary
    conditions conforming to observations.
\item
    Try to describe the modern acceleration stage, providing the ratio
    of the effective cosmological constant to the Planck density
    $\Lambdaef/m_4^4$ of the order $10^{-123}$.
\end{enumerate}

  We begin with the case of one factor space. Then \eqs (\ref{KE}) and
  (\ref{VE}) simplify to give
\bear
     S \eql \frac{\cV[d]}{2} \int d^{4}x\, \sqrt{\tg}\, (\sign F') L,
\nn
     L \eql \tR_4 +  K_E^{(1)}(\phi) (\d\phi)^2
                            - 2V_E^{(1)}(\phi),          \label{Lgen1}
\\
    K_E^{(1)}(\phi) \eql                                    \label{KE1}
        \frac{1}{4\phi^2} \biggl[
            6\phi^2 \biggl(\frac{F''}{F'}\biggr)^2\!
            -2 d_1 \phi \frac{F''}{F'}
\nnn \cm
        + \Half d_1 (d_1 + 2)\biggr]
                    + \frac {c_1 + c_2} {F' \phi},
\\
    V_E^{(1)}(\phi) \eql -\frac{\sign F'}{2F'^2}
    \biggl[ \frac{|\phi|} {d_1(d_1 {-} 1)}\biggr]^{d_1/2}
        \biggl[ F(\phi) + c_V \frac{\phi^2}{d_1} \biggr],
\nnn \cm
        c_V := c_1 +\frac{2c_2}{d_1-1}.                  \label{VE1}
\ear
  Here we take
\beq                                                     \label{F1}
    F = F(\phi) = \phi +c\phi^2 - 2\Lambda,\qquad  c,\ \Lambda = \const,
\eeq
  and, in accord with the definition (\ref{def-phi}), $\phi = d_1\phi_1$.

  In (\ref{Lgen1})--(\ref{VE1}) we have actually changed the sign of the
  Lagrangian in case $F' < 0$; to preserve the attractive nature of gravity
  for ordinary matter, the matter Lagrangian density should appear with an
  unusual sign from the beginning. As a result, the sign of the whole action
  of gravity and matter will be unusual, without any effect on the equations
  of motion; one can show that quantum transitions are then unaffected as
  well, see a discussion in \cite{VfromD}.

  The presence of the parameters $c_1$ and $c_2$ adds freedom in choosing
  the shape of the potential. The kinetic term also has a complex form which
  can significantly affect the field dynamics. An analysis of kinetic terms
  like (\ref{KE1}) of variable sign can lead to possibilities of interest,
  and we hope to return to this point in our future work.

  Let us employ the fact that chaotic inflation with a quadratic potential
  and the inflaton mass $m_\varphi \approx 10^{-6} m_4$ well conforms to
  the observational data. Therefore our task is simplified and reduced to
  finding such parameters $c,\ c_1$ and $c_2$ that the potential (\ref{VE1})
  near its minimum is approximated by a quadratic function with the above
  inflaton mass. It turns out to be possible with the following parameter
  values:
\bearr    \nq                                                \label{param1}
    d_1 = 4; \qquad
    c = 2.5 \cdot 10^4; \qquad
    c_1 + c_2 =0.6;
\nnn      \nq
    c_{\rm tot} := \frac{c_1}{d_1} + \frac{2c_2}{d_1(d_1-1)} =-0.62;
    \quad\ \Lambda = 0.2.
\ear

  With these parameter values, all basic requirements to inflation are
  satisfied. Thus, the duration of the inflationary period exceeds 60
  e-folds, the temperature fluctuations are $\sim 6\cdot 10^{-5}$,
  and the spectral index is $n_s = 0.943$, within observational bounds,
  $n_s = 0.958 \pm 0.016$ \cite{nasa}. Thus a single factor space is
  sufficient for obtaining a fairly good inflationary scenario.

  Since the constant $c$ has actually the dimension of length squared, it is
  $\sqrt{c} \sim 100$ that should be compared with the Planck length. So
  this model does not contain unnaturally large or small parameters.

  A serious shortcoming of this model is that it is unable to solve the
  problem of modern acceleration, including smallness of dark energy density.
  Indeed, it is easy to prove that slight variations of the parameters $c$,
  $c_1$ and $c_2$ could give rise to an arbitrarily small potential value at
  the minimum. However, though the values of these parameters are quite
  moderate, they should be extremely ``fine-tuned'' to fit the modern value
  of vacuum energy density. An attempt to solve this problem in a slightly
  more complex model is undertaken in the next section.

\section{Two factor spaces: inflation and modern acceleration}

\subsection*{Inflation}

  Additional opportunities emerge if the extra space is a product of two
  factor spaces, $\M_{d_1}\times \M_{d_2}$ of dimensions $d_1$ and $d_2$.
  For further analysis, let us make the situation more specific by putting
  $K_1 = K_2$, $d_1 = d_2$ and choosing the function
\beq                                        \label{F2}
        F(R) = R^2.
\eeq
  (Note that one of the coefficients in the initial Lagrangian can be chosen
  at will, e.g., equal to unity, without affecting the field equations; it
  simply specifies the scale for other coefficients.)

\begin{figure}
\includegraphics[scale=0.4]{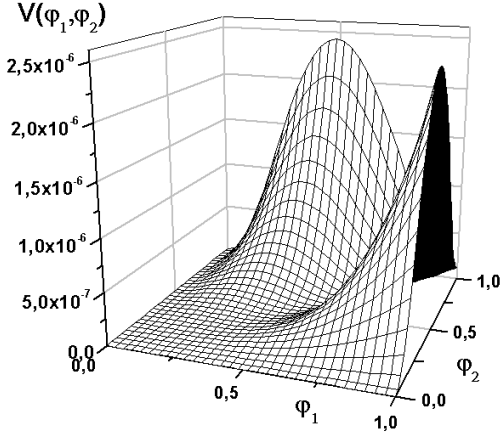}\\
\caption{Potential of the effective scalar fields for the model (\ref{S_D}),
    (\ref{F2}) with the parameter values given in (\ref{param2}).
    }
        \label{potential}
\end{figure}

  Fig.\,\ref{potential} presents the potential of the effective scalar fields
  for this model with the following choice of the parameter values:
\beq                                                        \label{param2}
    d_1 = d_2 = 5, \quad\
    c_V = -10.001,\quad\  c_1 + c_2 = 1.25\cdot 10^{3}.
\eeq
  All further numerical estimates will be obtained with these values. As
  follows from the above-said, at low energies (as compared to the Planck
  scale $\mD$) this model is equivalent to Einstein gravity with two scalar
  fields. In full similarity with Sec. \ref{single}, the constants $c_1$ and
  $c_2$ have actually the dimension of length squared, and their square
  roots are not unnaturally large or small.

  Note that, with the $c_V$ value chosen, a positive potential $V$
  (hence a positive effective cosmological constant) is obtained with $K_1
  =1$, i.e., spherical extra factor spaces. For other values of $c_V$,
  e.g., $c_V > 0$ we would need hyperbolic factor spaces.

  The inflationary period is characterized by moving down one of the steep
  slopes of the valley. The inflaton mass squared is proportional to the
  second-order derivative of the potential in the direction perpendicular to
  the valley (its bottom is located at $\phi_1 = \phi_2 = \phi_0$). It is
  this direction in which the field moves during inflation and oscillates
  during reheating at the post-inflationary stage. The specific value of
  $\phi_0$ depends on the initial value of the inflaton field at which the
  classical universe was born.

\begin{figure}
\includegraphics[scale=0.4]{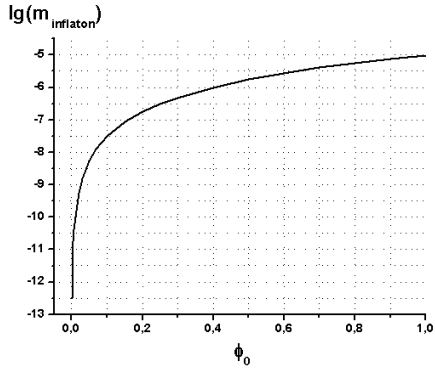}\\
\caption{Dependence of the effective inflaton mass (Planck units) on
the parameter $\phi_0$. }
            \label{inflaton}
\end{figure}

  Fig. \ref{inflaton} shows the dependence of the effective inflaton mass on
  the parameter $\phi_0$. In the framework of chaotic inflation, universes
  are created with different inflaton values under the horizon, leading to
  different values of $\phi_0$ and hence different inflaton masses. This is
  how this model solves the problem of smallness of the inflaton mass in
  Planck units.

  Post-inflationary particle production is a result of oscillations in the
  direction across the valley. The conditions suitable for our Universe
  correspond to the value $\phi_0 \simeq 0.5$. It is just such a value that,
  according to Fig.\,\ref{inflaton}, the inflaton mass, related to the
  second-order derivative of the potential in the direction across the
  valley, is $\sim 10^{-6} \sim 10^{13}$ GeV, which satisfactorily explains
  the observational data on the CMB temperature fluctuations.

\subsection*{Matter dominated stage}

  The inflationary stage ends with rapid field oscillations across the
  valley in Fig.\,1, on whose bottom, by our assumptions, $\phi_1 = \phi_2 =
  \phi/(2d_1)$.  These oscillations are accompanied by effective particle
  production in full agreement with the standard version of chaotic inflation
  with a quadratic potential. In the model under discussion, the energy
  density of the produced particles makes the material content of the
  Universe and affects not only the cosmological expansion rate but also the
  scalar field dynamics. The latter now corresponds to slow rolling down
  along the bottom of the potential valley.

  We assume a spatially flat cosmology in 4 dimensions, with the
  Einstein-frame metric $d{\wt s}{}^2_4 = dt^2 - a^2 (t) d\vec x{}^2$.
  So, with the choice (\ref{F2}), the action (\ref{SE}) leads to the
  Lagrangian
\bear                                                      \label{L2}
      L_E = R_4 + K (\phi) (d\phi)^2 - 2 V(\phi)
\ear 
and
\bearr
\label{K2a}
      K(\phi) (\d\phi)^2 = K_0 (\d\phi)^2/\phi^2 = 4K_0 (\d\beta)^2,
\nnnv \cm
      2K_0 = d_1^2 - d_1 + 3 + 4(c_1 + c_2),
\yyy                                                           \label{V2a}
      V (\phi) = V_0 |\phi|^{d_1} = V_1 \e^{-2d_1 \beta},
\nnn
      V_1 = -\frac{K_1}{8} \biggl(1 + \frac{c_V}{2d_1}\biggr), \quad\
      V_0 = V_1 [2 d_1 (d_1{-}1)]^{-d_1},
\nnn
\ear
  where $K_1 = \sign \phi = \sign F'(\phi)$ and $\beta (t) = \beta_1(t)$
  is, as before, the logarithm of the extra-dimension scale factor
  (which is in the present case the same for all extra dimensions),
  such that $d\phi/\phi = -2d\beta$, and $c_V$ has been defined in
  (\ref{VE1}). One can also notice that a usual form of the Lagrangian with a
  scalar field $\Phi$ and a potential $V_\Phi$ is obtained if we substitute
\[
      2\sqrt{K_0} \beta = \sqrt{8\pi G} \Phi, \cm  V_E = 8\pi G\, V_\Phi.
\]

  With (\ref{L2}), we can write two independent components of the
  Einstein-scalar equations for $\beta(t)$ and $a(t)$ as follows:
\bearr                                                         \label{H^2}
    3 H^2 = 2K_0 \dot\beta{}^2 + V_1\e^{-2d_1\beta} + 8\pi \rho_m
\yyy
    2K_0 \left[\ddot\beta                                      \label{b''}
              + 3 H \dot\beta \right] = d_1 V_1 \e^{-2d_1\beta},
\ear
  where $H = \dot a/a$ is the Hubble parameter.

  Let us begin with considering the matter dominated stage, which is the
  longest. The subsequent dark energy (DE) dominated stage will be discussed
  in the next subsection. The following simplifying assumptions will be used:
  (i) we neglect the pressure of matter, treating it as dust from the very
  beginning ($t=t_1$) thus ignoring a radiation-dominated stage; (ii) we
  neglect a possible direct interaction between matter and the scalar field;
  (iii) we neglect the scalar field contribution to the dynamics of $a(t)$ at
  the matter dominated stage $t_1 < t < t_2 \simeq 10^{10}$ years and, vice
  versa, we neglect the contribution of matter at the DE dominated stage $t
  > t_2$.

  So, neglecting the contribution of $\beta$ in (\ref{H^2}), we obtain for
  times $t_1 < t <  t_2$, as in the usual Big Bang scenario,
\beq                                       \label{HMD}
       H = 2/(3t)  \qquad {\rm at}\ \ t_1 < t < t_2.
\eeq
  To solve \eq (\ref{b''}) numerically, we take the following initial data
  corresponding to the end of the post-inflationary epoch:
\bearr
   \phi_1 (t_1) = \phi_2 (t_1) = \frac{\phi (t_1)}{2d_1} = 0.05;
    \qquad  \frac{d\phi}{dt} (t_1) = 0
\nnn
    \ \then \
    \e^{\beta (t_1)} = 4\sqrt{5}, \qquad \dot \beta(t_1) = 0.
\ear
  The initial time $t_1$ is chosen to be $t_1 = 9\ten {9}$ for
  definiteness.

  Numerical solution of \eq (\ref{b''}) then gives the following value of
  $\beta$ at $t = t_2$:
\bearr                                           \label{bt2}
    \e^{\beta (t_2)}=5.48255 \ten{11}\simeq 5.5 \ten{11}
\nnn \cm
                \ \then \       \phi(t_2)\simeq 1.3 \ten{-22}.
\ear
  This value of $\beta$ will be used in analyzing its dynamics at the modern
  stage for which the equations simplify and can be solved analytically.

\subsection*{Modern stage}

  The modern epoch $t > t_2$ is DE dominated. In the present approach, DE
  is represented by the scalar field $\phi$ (or equivalently $\beta$
  or $b = \e^{\beta}$) with the potential (\ref{V2a}), and the Universe
  dynamics is described by \eqs (\ref{H^2}), (\ref{b''}). In (\ref{H^2}) we
  now neglect the matter contribution.

  It is hard to solve this set of equations exactly. However, as the $\phi$
  field decreases (which corresponds to a growing size of the extra
  dimensions) along with a decreasing value of the potential (related to the
  effective cosmological constant), at some stage it becomes possible to
  treat this process as secondary slow rolling, for which the field dynamics
  is sufficiently simple and may be described analytically. Indeed, let us
  suppose
\beq                                                       \label{rolling}
       |\ddot \beta| \ll 3({\dot a}/a) \dot{\beta},
   \qquad
        K_0 \dot\beta{}^2 \ll 3 ({\dot a}/a)^2
\eeq
  and drop the corresponding terms in \eqs (\ref{H^2}) and (\ref{b''}).
  Then we can express $\dot a/a$ from (\ref{H^2}) and insert it to
  (\ref{b''}), getting
\beq
     d_1{\dot\beta} \e^{d_1\beta} = B_0
                    := \frac{d_1^2\sqrt{V_1}}{2\sqrt{3}K_0},
\eeq
  whence we find the evolution law for the extra-dimension scale factor
\beq
        \e^{\beta} = \Big[B_0(t-t_*) \Big]^{1/d_1},         \label{b(t)}
\eeq
  where $t_*$ is an integration constant
  ($t_* = t_2 - B_0^{-1}[b(t_2)]^{d_1}$). Substituting this result to
  (\ref{H^2}), we find the evolution law for $a(t)$:
\beq                                                        \label{a(t)}
        a(t) = a_* (t-t_*)^p, \cm p:= 2K_0/d_1^2,
\eeq
  where $a_*$ is an integration constant.

  With the parameters (\ref{param2}), some relevant constants are
\bearr
    V_1 = 1.25\ten{-4}, \cm 2K_0 = 5023,
\nnn
    p = \frac{5023}{25}\approx 201, \cm B_0 \approx 3.2 \ten{-5}.
\ear
  \eq (\ref{b(t)}) with the initial value (\ref{bt2}) gives
  the present size of the extra dimensions, at $t= t_0 = 13.7\ten {9}$ yr:
\beq                                                            \label{t*}
     b(t_0)=5.48259 \cdot10^{11} \simeq 5.5 \cdot10^{11}
     			\approx  9 \ten{-22}\,{\rm cm},
\eeq
  well within the observational limits. From (\ref{a(t)}) we find the Hubble
  constant $H_0= \dot a (t_0)/a (t_0)$ and the Hubble time $t_H = 1/H_0$:
\beq                                                     \label{H_0}
    H_0 \approx 1.25\ten{-61},  \qquad
    t_H \approx 8\ten {60} \approx 13.8\ten{9}\, {\rm yr},
\eeq
  in agreement with observations. The potential energy density $V$,
  coinciding with the DE density,
\beq
        V(\phi(t_0))\simeq 5.1 \ten{-123},              \label{V_0)}
\eeq
  also well agrees with observations.

  One can notice that in our model with $d_1 =5$ the function (\ref{b(t)})
  grows extremely slowly. The present value in (\ref{t*}) differs from that
  in (\ref{bt2}) only in the fifth decimal digit, so that the change is
  actually indistinguishable. The same is true for the DE density which thus
  behaves like a cosmological constant. The expansion law (\ref{a(t)}) with
  the exponent $p=201$ is really almost exponential, i.e., de Sitter, and the
  DE equation-of-state factor $w = p_{\rm DE}/\rho_{\rm DE}$ is very close
  to minus unity. Indeed, in the DE epoch, $ a(t)\sim t^{2/(3+3w)} $, hence
\[
        2/(3+3w) = 201\  \then \ w \approx -0.9967.
\]

  Lastly, one can verify that this solution fairly well satisfies the
  slow-rolling conditions (\ref{rolling}), which hold as long as $p \gg 1$.
  or, in terms of the input parameters of the theory, if $c_1 + c_2\gg d_1^2$.

  It is of interest that models of gravity (\ref{S_D}) where $F(R)$ contains
  a linear term do not lead to similar attractive results in the present
  approach.

\section{Conclusion}

  In the framework of pure curvature-nonlinear gravity with extra dimensions,
  it has been possible to describe (though in a rough approximation) the
  entire evolution of the Universe beginning with an inflationary stage and
  ending with the modern accelerated stage with sufficiently small dark
  energy density. In doing so, it has been possible to avoid unnaturally
  small or large parameter values in the initial Lagrangian. The small values
  of the inflaton mass and especially that of DE density agreeing with
  observations have been obtained from a Lagrangian whose dimensionless
  parameters differ from unity by no more than two orders of magnitude.

  Using a single extra factor space, it appears possible to explain the
  emergence of an inflaton, and choosing proper values of the parameter, it
  is possible to fulfil all requirements applicable to inflationary models
  and achieve an agreement with the observational data. However, to solve the
  problem of small DE density, it is necessary to invoke (at least) two extra
  factor spaces.

  The inflationary stage with an appropriate inflaton mass is again well
  described. Indeed, field fluctuations create universes with different
  initial field values. the potential in Fig.\,\ref{potential} (i.e., at
  fixed values of the initial Lagrangian parameters) has different
  curvatures at different points of the valley, which correspond to different
  inflaton masses. We live in a universe created by a suitable field
  fluctuation whose evolution leads to the observable inflaton mass.

  As to late-time evolution, it becomes possible to obtain in a natural way a
  small current value of the effective potential which plays the role of DE
  density (effective cosmological constant), $\Lambdaef \sim10^{-123}\
  m_4^4$). The form of our late-time solution shows that the size of the
  extra dimensions is slowly growing in the modern epoch. In the remote
  future, this size, which is so far invisible for modern instruments, is to
  grow to such values that will lead to drastic changes in the physical laws
  of our Universe. Let us stress, however, that such a model is only one
  particular opportunity contained in our approach. There are other models
  where the extra dimensions are stable at late times \cite{VfromD} making
  the effective physical constants also invariable.

  Our model with two factor spaces has the following advantages:
\begin{description}
\item [(a)]
    Its low-energy limit represents the Hilbert-Einstein action with
     appropriate accuracy.
\item [(b)]
    It describes inflation with an inflaton mass agreeing with observations;
\item [(c)]
    The size of the extra dimensions $b(t)$ never exceeded the experimental
    threshold $\sim 10^{-17}$ cm (though should exceed it in the remote
    future).
\item [(d)]
    At the modern stage, the scalar field density (actually, the potential
    $V(\phi)$ in proper units) describes the modern DE density
    $\sim 10^{-123} m_4^4$;
\item [(e)]
    The DE equation-of-state parameter $w$ satisfies the observational constraint $w<-0.8$.
\end{description}

  Since we have been working in the Einstein conformal frame, the problem of
  varying physical constants (above all, the effective Newtonian constant of
  gravity $G\eff$) did not emerge. One should note that even remaining in the
  Einstein frame, we could assume $\mD \ne m_4$, which would affect the
  estimated boundary between the classical and quantum worlds. In a more
  general framework, interpreting another conformal frame as the
  observational one (possibly but not necessarily the original Jordan frame),
  we would obtain a dependence of the constant $G\eff$ (hence the current
  Planck mass $m_4 = G\eff^{-1/2}$) on the size of extra dimensions, which
  in general can be not only time-dependent but also vary from point to point
  in space. In the cosmological context, models with variable $G\eff$ should
  not only satisfy the observational bounds on the variation rate
  $\dot G\eff/G\eff$ ($\lesssim 10^{-13}$ according to the recent tightest
  constraint \cite{muller07}) but also take into account the effect of $G(t)$
  on stellar evolution and processes in the early Universe. (Therefore,
  models with self-stabilizing extra dimensions like those discussed in
  \cite{VfromD, brost} can be more attractive.) In still more general models
  of this sort even the Planck constant $\hbar$ can be variable. A discussion
  of these problems is out of the scope of this paper and can be found, e.g.,
  in \cite{bm-predict, erice, Duff, Volovik, Rubin09jetp}.

\subsection*{Acknowledgment}

  We acknowledge financial support from RFBR grant 09-02-00677- .
  KB was also partly supported by NPK MU grant at PFUR.

\end{document}